\documentclass[preprint,superscriptaddress,nofootinbib]{revtex4}

  \usepackage{amssymb,amsmath,graphicx,subfigure,color}
  \usepackage[colorlinks]{hyperref}
  \begin{document}
  \title{Contributions from ${\Phi}_{B2}$ to the $B$ ${\to}$ $PP$ decays
    within the QCD factorization}
  \author{Qin Chang}
  \affiliation{Institute of Particle and Nuclear Physics,
              Henan Normal University, Xinxiang 453007, China}
  \author{Lili Chen}
  \affiliation{Institute of Particle and Nuclear Physics,
              Henan Normal University, Xinxiang 453007, China}
  \author{Yunyun Zhang}
  \affiliation{Institute of Particle and Nuclear Physics,
              Henan Normal University, Xinxiang 453007, China}
  \author{Junfeng Sun}
  \affiliation{Institute of Particle and Nuclear Physics,
              Henan Normal University, Xinxiang 453007, China}
  \author{Yueling Yang}
  \affiliation{Institute of Particle and Nuclear Physics,
              Henan Normal University, Xinxiang 453007, China}
  \date{\today}

  \begin{abstract}
  With the potential for the improvements of measurement precision,
  the refinement of theoretical calculation on hadronic $B$ weak
  decays is necessary. In this paper, we study the contributions of $B$ mesonic
  distribution amplitude ${\Phi}_{B2}$  within the QCD factorization
  approach, and find that ${\Phi}_{B2}$ contributes to only the
  nonfactorizable annihilation amplitudes for the $B$ ${\to}$ $PP$
  decays ($P$ denotes the ground $SU(3)$ pseudoscalar mesons).
  Although small, the ${\Phi}_{B2}$ contributions might be
  helpful for improving the performance of the QCD factorization
  approach, especially for the pure annihilation
  $B_{d}$ ${\to}$ $K^{+}K^{-}$ and $B_{s}$ ${\to}$ ${\pi}^{+}{\pi}^{-}$ decays.
  \end{abstract}
  \maketitle

  Because of successive impetus from both experiments and theoretical
  improvements, the study of nonleptonic $B$ meson weak decays has been
  one of the hot topics of particle physics.
  Most of the two-body hadronic $B$ decays with branching ratio larger
  than $10^{-6}$ have been investigated thoroughly and carefully
  at the {\sc BaBar} and Belle experiments \cite{pdg2018,epjc74.3026}
  in the past years.
  A huge amount of $B$ meson experimental data will be accumulated at the
  high luminosity colliders in the near future, about $50$ $ab^{-1}$ by
  the Belle-II detector at the $e^{+}e^{-}$ SuperKEKB collider \cite{1808.10567}
  and about $300$ $fb^{-1}$ by the LHCb Upgrade II detector at the hadron HL-LHC
  collider \cite{1808.08865,1812.07638}.
  With the advent of a new age of $B$ physics at the intensity frontier,
  besides some new phenomena, the unprecedented precision will offer
  a much more rigorous test on the standard model of elementary particles.
  The prospective experimental sensitivities for $B$ mesons require
  more and more accuracy of theoretical calculation.

  As is well known, the participation of the strong interactions make it
  very complicated to calculate the $B$ meson weak decays, especially for
  the nonleptonic cases.
  Based on power-counting rules in the heavy quark limits and perturbative
  QCD theory, some phenomenological models, such as QCD factorization (QCDF)
  \cite{prl83.1914,npb591.313,npb606.245,plb488.46,plb509.263,prd64.014036},
  perturbative QCD (pQCD) approach \cite{prd63.074006,plb504.6,prd63.054008,prd63.074009}
  and so on, have been developed and employed to compute the hadronic matrix
  elements (HMEs) describing the transformations between the initial $B$
  meson and final hadrons through local quark interactions.
  However, the nonperturbative contributions to HMEs bring theoretical
  results on branching ratios with many and large uncertainties, particularly
  for the internal $W$-boson emission and the neutral current processes.
  To reduce theoretical uncertainties and satisfy the precision requirements
  of experimental analysis, a careful and comprehensive examination
  of all possible nonperturbative factors within a phenomenological model is
  necessary. In this paper, the contributions from the $B$ meson wave functions
  will be reassessed in detail within the theoretical framework of QCDF.

  Wave functions (WFs) or distribution amplitudes (DAs) of the $B$ meson are
  the essential ingredients of the master  formulas in QCDF \cite{npb591.313} and
  pQCD \cite{plb504.6} approaches to evaluate the nonfactorizable  contributions
  to HMEs, such as the spectator scattering amplitudes.
  However, the knowledge of the $B$ mesonic WFs and DAs is still limited so far.
  It is intuitive that the component quarks of a hadron should move with the
  same velocity to form a color singlet, and thus the valence quarks would
  share momentum fractions according to their masses.
  It is expected that the $B$ mesonic DAs should be very asymmetric with
  ${\xi}$ at the scales of order $m_{b}$ or smaller, if the light spectator
  quark carries a longitudinal momentum fraction ${\xi}$ ${\sim}$
  ${\cal O}({\Lambda}_{\rm QCD}/m_{b})$, where ${\Lambda}_{\rm QCD}$ and $m_{b}$
  are respectively the characteristic QCD scale and the mass of $b$ quark.
  Generally, the $B$ meson is described by two scalar functions
  up to the leading power in $1/m_{b}$ \cite{prd55.272,npb592.3,npb625.239,npb642.263},
  which is written as \cite{npb591.313}
  \begin{equation}
  {\langle}0{\vert}\bar{q}_{\alpha}(z)[...]b_{\beta}(0){\vert}
   \bar{B}(p){\rangle}\, =\, -\frac{i\,f_{B}}{4}\, \big\{
   \big( \!\!\not{p}+m_{b} \big)\, {\gamma}_{5} \big\}_{{\beta}{\gamma}}{\int}d{\xi}\,
  e^{-i{\xi}p_{+}z_{-}} \Big[ {\Phi}_{B1}({\xi})
  +\not{n}_{-}{\Phi}_{B2}({\xi})\Big]_{{\gamma}{\alpha}}
  \label{b-meson-wfs-01},
  \end{equation}
  where the dots denote the path-ordered exponential gauge factor;
  the light spectator quark moves along the light-like $z_{-}$ line;
  $n_{-}$ $=$ $(1,0,0,-1)$ is a null vector; and the normalization
  conditions of DAs are \cite{npb591.313}
  \begin{equation}
  {\int}_{0}^{1}d{\xi}\,{\Phi}_{B1}({\xi})\, =\, 1
  \label{b-meson-wfs-02},
  \end{equation}
  \begin{equation}
  {\int}_{0}^{1}d{\xi}\,{\Phi}_{B2}({\xi})\, =\, 0
  \label{b-meson-wfs-03}.
  \end{equation}

  According to the conventions of Refs.\cite{prd55.272,npb592.3},
  ${\Phi}_{B1}$ $=$ ${\phi}_{B}^{+}$ and
  ${\Phi}_{B2}$ $=$ $({\phi}_{B}^{+}$ $-$ ${\phi}_{B}^{-})/2$.
  Generally, the two functions ${\phi}_{B}^{\pm}$
  are not identical, ${\phi}_{B}^{+}$ ${\neq}$ ${\phi}_{B}^{-}$,
  and satisfy the relation ${\phi}_{B}^{+}({\xi})$ $+$
  ${\xi}\,{\phi}_{B}^{-{\prime}}({\xi})$ $=$ $0$ \cite{npb592.3}.
  So, ${\Phi}_{B2}$ ${\neq}$ $0$.
  The contributions of ${\Phi}_{B2}$ part are suppressed by the
  power factor of ${\Lambda}_{\rm QCD}/m_{b}$, compared with those of
  ${\Phi}_{B1}$.
  In the actual calculations for the $B$ ${\to}$ $PP$ decays with the QCDF
  approach ($P$ denotes the light $SU(3)$ ground pseudoscalar meson),
  for example in Ref.\cite{npb606.245}, only the contributions from
  ${\Phi}_{B1}$ part are considered appropriately, while those from
  ${\Phi}_{B2}$ part are not included explicitly.
  It should be pointed out that the value of ${\Lambda}_{\rm QCD}/m_{b}$
  is not a negligible number, because the mass of the $b$ quark is finite
  rather than infinite.
  It has been shown in Refs.\cite{npb625.239,npb642.263,
  prd71.034018,epjc28.515,prd74.014027} that there is a large contribution of
  ${\Phi}_{B2}$ to the hadronic $B$ ${\to}$ ${\pi}$ transition formfactors
  within the pQCD approach, and its share could reach up to ${\sim}$ $30\%$
  with some specific inputs \cite{epjc28.515,prd74.014027}.
  This means that the contributions of ${\Phi}_{B2}$ to branching ratios
  for the $W$ emission processes can reach up to ${\sim}$ $70\%$ for
  some cases.
  The ${\Phi}_{B2}$ contribution that were neglected in most
  cases should be given due attention with the QCDF approach, which is
  the focus of this paper.

  Here, it should be pointed out that a possibly large
  contribution of ${\Phi}_{B2}$ to formfactors is present only with the
  pQCD approach rather than the QCDF approach, due to different
  understandings on the nature of the hadronic transition formfactors.
  With the pQCD approach \cite{prd63.074006,plb504.6,prd63.054008,prd63.074009},
  it is assumed that the light quark with a soft momentum of
  ${\cal O}({\Lambda}_{\rm QCD})$ in the initial $B$ meson should
  interact with a hard gluon, so it could receive a large boost in order
  to form a colorless final state with a light energetic quark originating
  from the $b$ quark decaying interaction point. It is therefore arguable
  that the hadronic transition formfactors are computable perturbatively
  with the help of the Sudakov factor regulation on soft contributions.
  The hadronic transition formfactors are written as the convolution
  of wave functions of both the $B$ meson and final hadron.
  Contrarily, it is argued \cite{npb591.313,npb592.3} with the QCDF approach
  that the hard and soft contributions to the heavy-to-light formfactors
  have the same scaling behavior, and the hard contributions are suppressed
  by one power of ${\alpha}_{s}$ compared with the soft contributions.
  Because of the dominance of soft contributions, the formfactors for
  the transition between $B$ meson and light hadron are not fully calculable
  with the perturbative QCD theory. So, the formfactors are regarded
  as nonperturbative inputs with the QCDF approach, and therefore have nothing
  to do with the $B$ mesonic wave functions.

  We will concentrate on the $B$ ${\to}$ $PP$ decays for the moment.
  Up to power corrections of $1/m_{b}$, the general QCDF formula
  of HMEs for an effective operator $\hat{O}_{i}$ is written
  as \cite{npb591.313},
  \begin{eqnarray}
  {\langle}P_{1}P_{2}{\vert}\hat{O}_{i}{\vert}\bar{B}{\rangle} &=&
   F_{0}^{B{\to}P_{1}}{\int}_{0}^{1}dx\,T^{I}_{i}(y)\,{\phi}_{P_{2}}(x)
  +F_{0}^{B{\to}P_{2}}{\int}_{0}^{1}dy\,H^{I}_{i}(x)\,{\phi}_{P_{1}}(y)
   \nonumber \\ &+&
  {\int}_{0}^{1}d{\xi}\,dx\,dy\,T^{II}_{i}({\xi},x,y)\,
  {\phi}_{B}({\xi})\,{\phi}_{P_{1}}(y)\,{\phi}_{P_{2}}(x)
  \label{qcdf-01},
  \end{eqnarray}
  where $F_{0}^{B{\to}P_{i}}$ denotes the formfactor; $T^{I}$, $H^{I}$
  and $T^{II}$ are hard scattering kernels; the mesonic DAs,
  ${\phi}_{P_{2}}(x)$ and ${\phi}_{P_{1}}(y)$, are the functions of
  longitudinal momentum fractions $x$ and $y$ of light quarks.

  For the two terms in the first line of Eq.(\ref{qcdf-01}),
  soft contributions are assumed to be embodied in the formfactors
  $F_{0}^{B{\to}P_{i}}$ and DAs.
  Contributions of $T^{I}$ and $H^{I}$ are dominated by hard gluon exchange.
  So these contributions, which are irrelevant to $B$ mesonic wave functions,
  are considered as perturbative corrections to the naive factorization
  formula, which involve only decay constants and formfactors, but no DAs.

  \begin{figure}[t]
  \subfigure[]{\includegraphics[scale=1]{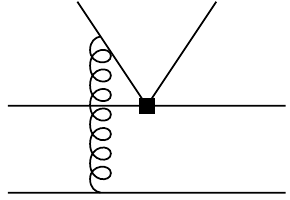}}\qquad\qquad
  \subfigure[]{\includegraphics[scale=1]{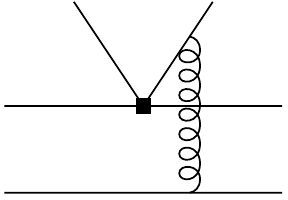}}
  \caption{The spectator scattering interactions.}
  \label{fig-hs}
  \end{figure}

  The term in the second line of Eq.(\ref{qcdf-01}) corresponds to
  nonfactorizable contributions. The
  spectator scattering interactions (see Fig.\ref{fig-hs})
  entangle the initial $B$ meson
  with the final hadrons, which make separating one hadron from others
  impossible. Therefore, the spectator scattering amplitudes are
  usually written as the convolution integral of the hard kernels
  $T^{II}$ and all participating DAs. The
  hard spectator scattering amplitudes contain the contributions
  from both ${\Phi}_{B1}$ and ${\Phi}_{B2}$, and can be written as
  \begin{eqnarray}
  H_{k}(P_{1},P_{2})\, =\, H_{k}^{B1}(P_{1},P_{2})+H_{k}^{B2}(P_{1},P_{2})
  \label{qcdf-hs-01},
  \end{eqnarray}
  where $P_{1}$ is the emitted meson; $P_{2}$ is the recoiled meson
  that incorporates the spectator quark from $B$ meson into itself;
  $H_{k}^{B1}$ ($H_{k}^{B2}$) is the contribution from ${\Phi}_{B1}$
  (${\Phi}_{B2}$); the subscript $k$ on $H_{k}$ refers to the possible
  Dirac current structure ${\Gamma}{\otimes}{\Gamma}$ of an operator $\hat{O}$,
  namely, $k$ $=$ $1$, $2$ and $3$ correspond to ${\Gamma}{\otimes}{\Gamma}$
  $=$ $(V-A){\otimes}(V-A)$, $(V-A){\otimes}(V+A)$ and $-2(S-P){\otimes}(S+P)$
  respectively.
  After the straightforward calculation, we find that considering the $SU(3)$
  flavor symmetry, the expressions of $H_{1}^{B1}$ and $H_{2}^{B1}$ are entirely
  consistent with Eq.(47) and Eq.(48) of Ref.\cite{npb675.333},
  and $H_{3}^{B1}$ $=$ $0$. Our calculations also show that
  $H_{k}^{B2}$ corresponding to Fig.\ref{fig-hs}(a) and Fig.\ref{fig-hs}(b)
  are nonzero.
  Moreover, the terms of both ${\int}_{0}^{1}\frac{{\Phi}_{P1}(y)}{\bar{y}^{2}}dy$
  and ${\int}_{0}^{1}\frac{{\Phi}_{P1}^{p}(y)}{\bar{y}^{2}}dy$
  appear in $H_{k}^{B2}$, where ${\Phi}_{P1}(y)$ and ${\Phi}_{P1}^{p}(y)$
  are the leading twist (twist-2) and twist-3 DAs of the emitted meson $P_{1}$
  and $\bar{y}$ $=$ $1$ $-$ $y$.
  It is clearly seen that with the asymptotic forms of
  ${\Phi}_{P1}(y)$ $=$ $6\,y\,\bar{y}$ and ${\Phi}_{P1}^{p}(y)$ $=$ $1$,
  the integrals of ${\int}_{0}^{1}\frac{{\Phi}_{P1}(y)}{\bar{y}^{2}}dy$
  and  ${\int}_{0}^{1}\frac{{\Phi}_{P1}^{p}(y)}{\bar{y}^{2}}dy$ exhibit
  logarithmic and linear infrared divergences.
  Fortunately, because of the opposite sign between the emitted quark and
  antiquark propagators plus the condition of Eq.(\ref{b-meson-wfs-03}),
  the contributions of $H_{k}^{B2}$ exactly cancel each other out.
  The total contributions from ${\Phi}_{B2}$ to spectator scattering
  amplitudes are zero.

  \begin{figure}[t]
  \subfigure[]{\includegraphics[scale=1]{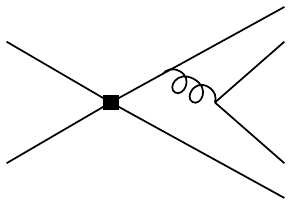}}\qquad
  \subfigure[]{\includegraphics[scale=1]{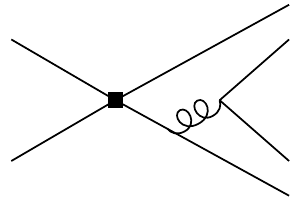}}\qquad
  \subfigure[]{\includegraphics[scale=1]{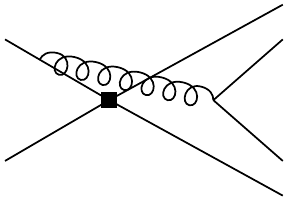}}\qquad
  \subfigure[]{\includegraphics[scale=1]{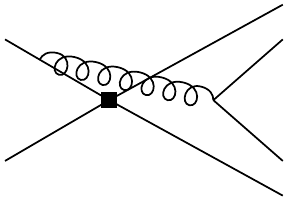}}
  \caption{The weak annihilation interactions, where (a) and (b) are
  factorizable diagrams, (c) and (d) are nonfactorizable diagrams.}
  \label{fig-ann}
  \end{figure}

  Compared with the leading contributions, the weak annihilation (WA)
  contributions are thought to be suppressed by one power of
  ${\Lambda}_{\rm QCD}/m_{b}$ \cite{npb591.313}.
  However, the WA contributions are significant and can not be ignored
  in practical application of the QCDF approach to the hadronic $B$ decays
  \cite{npb606.245,npb675.333,prd65.074001,prd65.094025,npb774.64}.
  Therefore, the QCDF master formula of Eq.(\ref{qcdf-01}) is generalized
  to estimate the WA contributions.
  The WA interactions have two types of topologies within the QCDF approach.
  The nonfactorizable and factorizable topologies respectively correspond
  to gluon emission from the initial $B$ meson and final quarks,
  see Fig.\ref{fig-ann}.
  The factorizable WA amplitudes can be written as the product of the
  time-like $0$ ${\to}$ $P_{1}P_{2}$ formfactors and the integral
  of $B$ mesonic WFs, see Fig.\ref{fig-ann}(a) and (b).
  With the normalization condition of Eq.(\ref{b-meson-wfs-03}), it is
  clearly seen that ${\Phi}_{B2}$ contributes nothing to the factorizable
  WA amplitudes $A^{f}_{k}$, where the superscript $f$ means factorizable,
  {\em i.e.}, gluon emission from the final quarks; the subscript $k$
  has the same meaning as that of $H_{k}$ in Eq.(\ref{qcdf-hs-01}).
  The nonfactorizable WA amplitudes, corresponding to Fig.\ref{fig-ann}(c)
  and (d), can be written as the convolution integral of all participating
  hadronic DAs, and contain the contributions from both ${\Phi}_{B1}$ and
  ${\Phi}_{B2}$.
  \begin{eqnarray}
  A^{i}_{k}\, =\, A^{i,B1}_{k}+A^{i,B2}_{k}
  \label{qcdf-wa-01},
  \end{eqnarray}
  where the superscript $i$ means gluon emission from the initial $B$ meson;
  $A^{i,B1}_{k}$ ($A^{i,B2}_{k}$) is the contribution from ${\Phi}_{B1}$
  (${\Phi}_{B2}$).
  The expressions of $A^{i,B1}_{k}$ have been explicitly given by
  Eq.(62) of Ref.\cite{npb606.245} and Eq.(54) of Ref.\cite{npb675.333}.
  Here, we will give the new components $A^{i,B2}_{k}$.
  \begin{eqnarray}
  A^{B2}\, =\, {\pi}\,{\alpha}_{s}\, {\int}_{0}^{1}{\xi}\,{\Phi}_{B2}({\xi})\,d{\xi}
        \, =\, {\pi}\,{\alpha}_{s}\, {\langle}{\xi}{\rangle}_{B2}
  \label{qcdf-wa-02},
  \end{eqnarray}
  \begin{eqnarray}
  A^{i,B2}_{1}\, =\, -A^{B2}\, {\int}_{0}^{1}\frac{dx}{\bar{x}}
  {\int}_{0}^{1}\frac{dy}{y} \Big\{
  2\,\frac{{\Phi}_{P2}(x)\,{\Phi}_{P1}(y)}{1-x\,\bar{y}}
  -r_{\chi}^{P1}\,r_{\chi}^{P2}\, \frac{\bar{x}}{y}\,
  \frac{{\Phi}_{P2}^{p}(x)\,{\Phi}_{P1}^{p}(y)}{1-x\,\bar{y}} \Big\}
  \label{qcdf-wa-03},
  \end{eqnarray}
  \begin{eqnarray}
  A^{i,B2}_{2}\, =\, A^{B2}\, {\int}_{0}^{1}\frac{dx}{\bar{x}}
  {\int}_{0}^{1}\frac{dy}{y} \Big\{
  2\,\frac{{\Phi}_{P2}(x)\,{\Phi}_{P1}(y)}{\bar{x}\,y}
  -r_{\chi}^{P1}\,r_{\chi}^{P2}\, {\Phi}_{P2}^{p}(x)\,{\Phi}_{P1}^{p}(y)\,
   \Big[ \frac{\bar{x}}{1-x\,\bar{y}}-\frac{x}{\bar{x}\,y} \Big] \Big\}
  \label{qcdf-wa-04},
  \end{eqnarray}
  \begin{eqnarray}
  A^{i,B2}_{3}\, =\, -A^{B2}\, {\int}_{0}^{1}\frac{dx}{\bar{x}}
  {\int}_{0}^{1}\frac{dy}{y} \Big\{ 2\,r_{\chi}^{P2}\,
   \frac{x\,{\Phi}_{P2}^{p}(x)\,{\Phi}_{P1}(y)}{1-x\,\bar{y}}
  + r_{\chi}^{P1}\,{\Phi}_{P2}(x)\,{\Phi}_{P1}^{p}(y)\,
   \Big[ \frac{y-\bar{y}}{1-x\,\bar{y}}+\frac{1}{\bar{x}\,y} \Big] \Big\}
  \label{qcdf-wa-05},
  \end{eqnarray}
  where the factor $r_{\chi}^{P}$ $=$
  $\frac{2\,m_{P}^{2}}{\bar{m}_{b}\,(\bar{m}_{q_{1}}+\bar{m}_{q_{2}})}$.

  It is easy to find that contributions from ${\Phi}_{B2}$ to the WA amplitudes
  are nonzero, because the moment parameter ${\langle}{\xi}{\rangle}_{B2}$ is
  nonzero. Hence, ${\Phi}_{B2}$ may present nontrivial effects on the observables
  of hadronic $B$ decays, especially for the WA dominant ones.

  In order to better investigate the ${\Phi}_{B2}$ contributions and eliminate
  other pollution, the pure WA decays $B_{d}$ ${\to}$ $K^{+}K^{-}$ and
  $B_{s}$ ${\to}$ ${\pi}^{+}{\pi}^{-}$ will be restudied in this paper.
  Although their branching ratios are tiny, they have been measured
  accurately by now \cite{hfag}.
   \begin{equation}
   {\cal B}(B_{s}{\to}{\pi}^{+}{\pi}^{-})\,=\,(6.7{\pm}0.8){\times}10^{-7}
   \label{exp-bs-pipi},
   \end{equation}
   \begin{equation}
   {\cal B}(B_{d}{\to}K^{+}K^{-})\,=\,(8.0{\pm}1.5){\times}10^{-8}
   \label{exp-bd-kk}.
   \end{equation}

  With the asymptotic twist-2 and -3 DAs, ${\Phi}_{P}(u)$ $=$ $6\,u\,\bar{u}$ and
  ${\Phi}_{P}^{p}(u)$ $=$ $1$,
  the integrals in Eq.(\ref{qcdf-wa-03}-\ref{qcdf-wa-05}) exhibit
  logarithmic and linear infrared divergences.
  For an estimation of the WA contributions from ${\Phi}_{B2}$, these divergent endpoint
  integrals will be parameterized by the commonly used notations within
  the QCDF approach \cite{npb606.245,npb675.333,npb774.64}.
  \begin{equation}
  {\int}_{0}^{1} \frac{du}{u}\ {\to}\ X_{A}
  \label{xa-01},
  \end{equation}
  \begin{equation}
  {\int}_{0}^{1} \frac{du}{u^{2}}\ {\to}\ X_{L}
  \label{xa-02},
  \end{equation}
  \begin{equation}
  {\int}_{0}^{1} du\,\frac{{\ln}u}{u}\ {\to}\ -\frac{1}{2}\,X_{A}^{2}
  \label{xa-03},
  \end{equation}
  \begin{equation}
  {\int}_{0}^{1} du\,\frac{{\ln}u}{ u^{2} }\ {\to}\ X_{L}-X_{A}-X_{L}\,X_{A}
  \label{xa-04}.
  \end{equation}
  The phenomenological parameters $X_{A}$ and $X_{L}$, are usually
  treated as universal for hadronic $B$ decays in previous literatures
  \cite{npb606.245,npb675.333,prd65.074001,prd65.094025,npb774.64}.

  With the above parameterization scheme, the WA amplitudes can
  be rewritten as
   \begin{equation}
   {A}_{1}^{i,B2}\, =\, -A^{B2}\,\Big\{
     12\,({\pi}^{2}-6)-r_{\chi}^{P1}\,r_{\chi}^{P2}\,\Big[
     \frac{{\pi}^{2}}{6}+\frac{1}{2}\,X_{A}^{2}+X_{L}X_{A}+X_{A}-X_{L}\Big]\Big\}
  \label{qcdf-p-wa-01},
  \end{equation}
   \begin{equation}
   {A}_{2}^{i,B2}\, =\, A^{B2}\,\Big\{
     72\,(X_{A}-1)^{2}-r_{\chi}^{P1}\,r_{\chi}^{P2}\,\Big[
     \frac{{\pi}^{2}}{6}+\frac{1}{2}\,X_{A}^{2}+X_{L}X_{A}-X_{L}^{2}\Big]\Big\}
  \label{qcdf-p-wa-02},
  \end{equation}
   \begin{equation}
   {A}_{3}^{i,B2}\, =\, -A^{B2}\,6\,\Big\{
     \Big[ \frac{{\pi}^{2}}{6}+\frac{1}{2}\,X_{A}^{2}-X_{A}\Big]
     (2\,r_{\chi}^{P2}-r_{\chi}^{P1})+r_{\chi}^{P1}\,
     (X_{A}X_{L}-X_{L}+1) \Big\}
  \label{qcdf-p-wa-03}.
  \end{equation}

  The parameters of $X_{A}$ and $X_{L}$ including part of
  strong phases are complex, and are usually parameterized as
  \cite{npb606.245,npb675.333,prd65.074001,prd65.094025,npb774.64}
   \begin{equation}
   X_{A}\, =\, (1+{\rho}_{A}\,e^{i{\phi}_{A}})\,
     {\ln}\frac{m_{B}}{{\Lambda}_{h}}
   \label{xa-01},
   \end{equation}
   \begin{equation}
   X_{L}\, =\, (1+{\rho}_{A}\,e^{i{\phi}_{A}})\,
     \frac{m_{B}}{{\Lambda}_{h}}
   \label{xa-02},
   \end{equation}
  where ${\Lambda}_{h}$ $=$ $0.5$ GeV \cite{npb606.245,npb675.333},
  and ${\phi}_{A}$ is an undetermined strong phase.
  In addition, according to the relations given by Refs.\cite{prd55.272,npb592.3},
  the moment parameter in Eq.(\ref{qcdf-wa-02}) is
   \begin{equation}
   {\langle}{\xi}{\rangle}_{B2}
   \, =\, \frac{1}{2}\,\Big({\langle}{\xi}{\rangle}_{+}-{\langle}{\xi}{\rangle}_{-}\Big)
   \, =\, \frac{\bar{\Lambda}}{3\,m_{b}}\,
   \label{xa-03},
   \end{equation}
  with ${\langle}{\xi}{\rangle}_{+}$ $=$ $2\,{\langle}{\xi}{\rangle}_{-}$
  $=$ $\frac{4}{3}\,\frac{\bar{\Lambda}}{m_{b}}$ and
  $\bar{\Lambda}$ $=$ $m_{B}$ $-$ $m_{b}$ ${\approx}$
  $0.55$ GeV \cite{prd55.272}.
  Using the exponential type model for $B$ meson DAs
   \begin{equation}
  {\phi}_{B_{q}}^{+}({\xi})\,=\, N^{+}\,{\xi}\,
  {\exp} \Big(-\frac{ {\xi}\,m_{B_{q}} }{ {\omega}_{B_{q}} } \Big)
   \label{gn-da-01},
   \end{equation}
   \begin{equation}
  {\phi}_{B_{q}}^{-}({\xi})\,=\, N^{-}\,
  {\exp} \Big(-\frac{ {\xi}\,m_{B_{q}} }{ {\omega}_{B_{q}} } \Big)
   \label{gn-da-02},
   \end{equation}
  where $N^{\pm}$ is the normalization constant determined via
  ${\int}_{0}^{1}{\phi}_{B_{q}}^{\pm}({\xi})d{\xi}$ $=$ $1$,
  one can obtain ${\langle}{\xi}{\rangle}_{B2}$ $=$
  $0.042{\pm}0.01$ with the shape parameter ${\omega}_{B_{s}}$
  $=$ $0.45{\pm}0.10$ GeV for $B_{s}$ meson \cite{epjc73.2437}, and
  ${\langle}{\xi}{\rangle}_{B2}$ $=$ $0.039{\pm}0.01$ with
  ${\omega}_{B_{d}}$ $=$ $0.42{\pm}0.10$ GeV for
  $B_{d}$ meson \cite{prd74.014027}, which are basically in agreement
  with the estimation of Eq.(\ref{xa-03}).

  Using the commonly used notations in the QCDF approach
  \cite{npb606.245,npb675.333,prd65.074001,prd65.094025,npb774.64},
  the amplitudes for the pure WA decays $B_{d}$ ${\to}$ $K^{+}K^{-}$
  and $B_{s}$ ${\to}$ ${\pi}^{+}{\pi}^{-}$ are written as
   \begin{equation}
  {\cal A}(B_{s}{\to}{\pi}^{+}{\pi}^{-})\,=\,
    i\,\frac{G_{F}}{\sqrt{2}}\,f_{B_{s}}\,f_{\pi}^{2}
   \Big\{ V_{ub}^{\ast}V_{us}\,\Big( b_{1}+2\,b_{4}+\frac{1}{2}\,b_{4,{\rm EW}}\Big)
   +V_{cb}^{\ast} V_{cs}\,\Big(2\,b_{4}+\frac{1}{2}\,b_{4,{\rm EW}} \Big)\Big\}
   \label{bs-pipi-01},
   \end{equation}
   \begin{equation}
  {\cal A}(B_{d}{\to}K^{+}K^{-})\,=\,
  i\,\frac{G_{F}}{\sqrt{2}}\,f_{B_{d}}\,f_{K}^{2}
   \Big\{ V_{ub}^{\ast}V_{ud}\,\Big( b_{1}+2\,b_{4}+\frac{1}{2}\,b_{4,{\rm EW}}\Big)
   +V_{cb}^{\ast} V_{cd}\,\Big(2\,b_{4}+\frac{1}{2}\,b_{4,{\rm EW}} \Big)\Big\}
   \label{bd-kk-01},
   \end{equation}
  where the Fermi weak coupling constant $G_{F}$ ${\simeq}$ $1.166{\times}10^{-5}$
  ${\rm GeV}^{-2}$ \cite{pdg2018}; $f_{B_{q}}$, $f_{\pi}$ and $f_{K}$ are
  decay constants; $V_{ij}$ ($i$ $=$ $u$, $c$ and $j$ $=$ $d$, $s$, $b$)
  is the Cabibbo-Kobayashi-Maskawa (CKM) matrix element.
  The definition of parameter $b_{i}$ is
   \begin{equation}
   b_{1}\, =\, \frac{C_{F}}{N_{c}^{2}}\,C_{1}\, A_{1}^{i}
   \label{qcdf-coe-b1},
   \end{equation}
   \begin{equation}
   b_{4}\, =\, \frac{C_{F}}{N_{c}^{2}}\, \Big[
   C_{4}\, A_{1}^{i} + C_{6}\, A_{2}^{i} \Big]
   \label{qcdf-coe-b4},
   \end{equation}
   \begin{equation}
   b_{4,{\rm EW}}\, =\, \frac{C_{F}}{N_{c}^{2}}\, \Big[
   C_{10}\, A_{1}^{i} + C_{8}\, A_{2}^{i} \Big]
   \label{qcdf-coe-b4ew},
   \end{equation}
  where $C_{F}$ $=$ $4/3$ is the color factor; $N_{c}$ $=$ $3$
  is the number of colors; $C_{i}$ is the Wilson coefficient;
  $A^{i}_{k}$ is the amplitude building block of Eq.(\ref{qcdf-wa-01}).

  To provide a quantitative estimate of the ${\Phi}_{B2}$ contributions,
  the inputs listed in Table.\ref{tab:inputs} are used in our numerical
  calculation. Their central values will be regarded as the default
  inputs unless otherwise specified.

   \begin{table}[t]
   \caption{The input parameters \cite{pdg2018}.}
   \label{tab:inputs}
   \begin{ruledtabular}
   \begin{tabular}{lll}
    $m_{b}$ $=$ $4.78{\pm}0.06$ GeV, &
    $\bar{m}_{b}(\bar{m}_{b})$ $=$ $4.18^{+0.04}_{-0.03}$ GeV, &
    $\bar{m}_{s}(2\,{\rm GeV})$ $=$ $95{\pm}5$ MeV, \\
    $\frac{\bar{m}_{s}(2\,{\rm GeV})}{\bar{m}_{u,d}(2\,{\rm GeV})}$ $=$ $27.3{\pm}0.7$, &
    $f_{\pi}$ $=$ $130.2{\pm}1.7$ MeV, &
    $f_{K}$ $=$ $155.6{\pm}0.4$ MeV, \\
    $m_{B_{d}}$ $=$ $5279.63{\pm}0.15$ MeV, &
    $f_{B_{d}}$ $=$ $187.1{\pm}4.2$ MeV, &
    ${\tau}_{B_{d}}$ $=$ $1.520{\pm}0.004$ ps, \\
    $m_{B_{s}}$ $=$ $5366.89{\pm}0.19$ MeV, &
    $f_{B_{s}}$ $=$ $227.2{\pm}3.4$ MeV, &
    ${\tau}_{B_{s}}$ $=$ $1.509{\pm}0.004$ ps.
   \end{tabular}
   \end{ruledtabular}
   \end{table}
  \begin{figure}[t]
  \subfigure[]{ \includegraphics[width=0.4\textwidth]{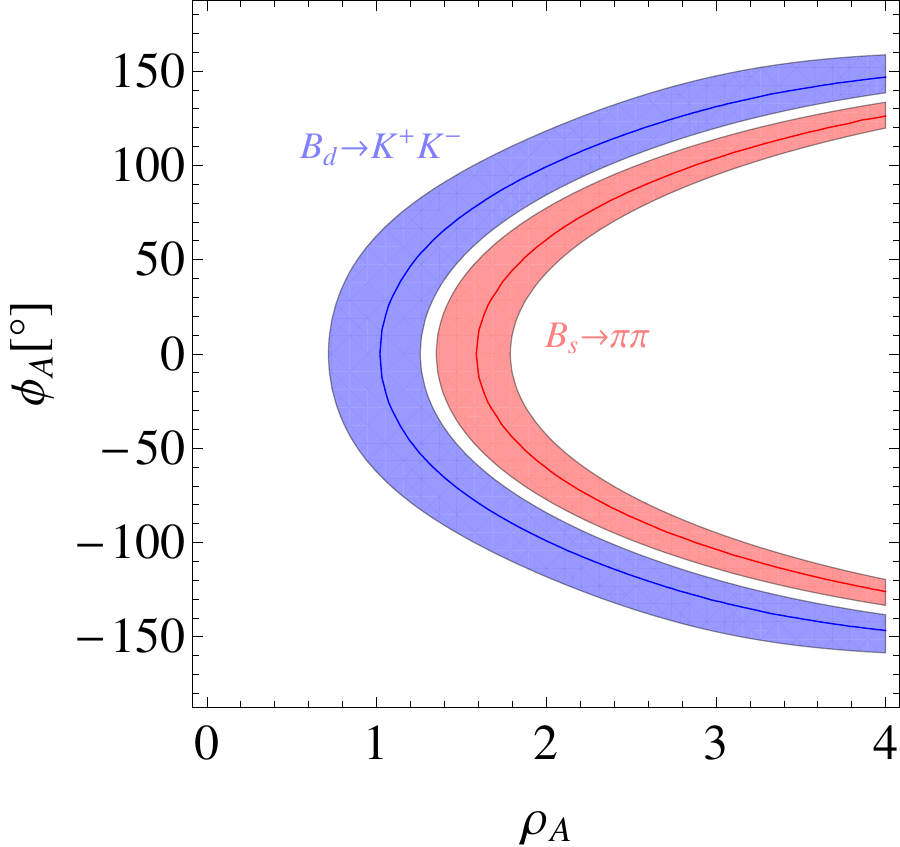} } \qquad
  \subfigure[]{ \includegraphics[width=0.4\textwidth]{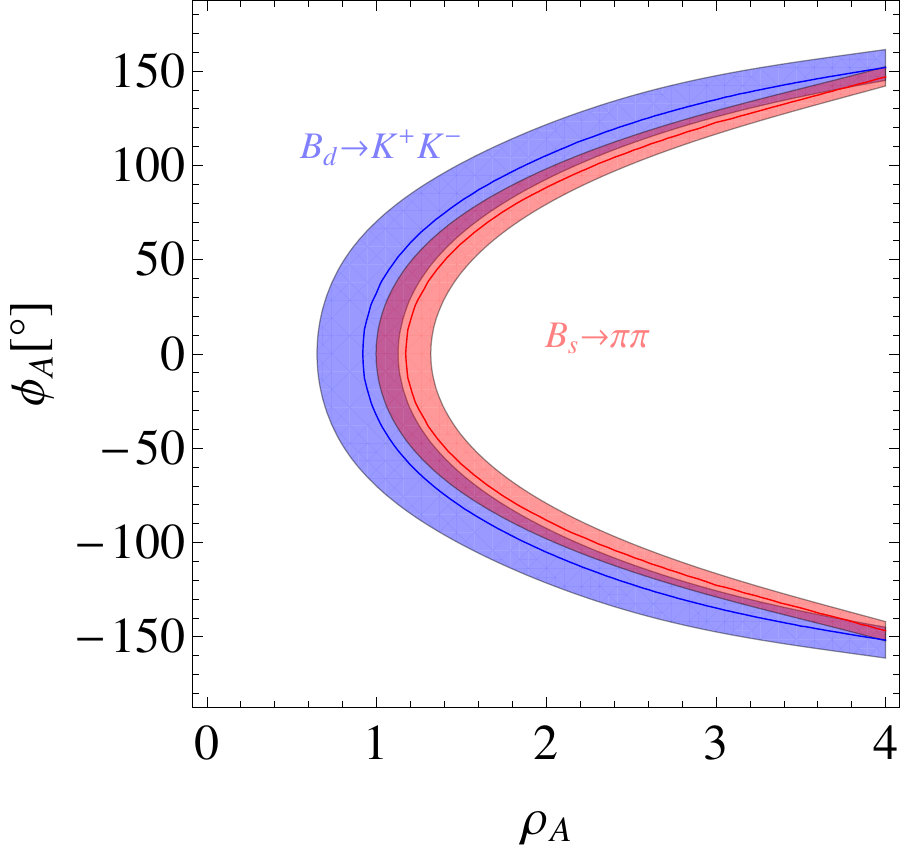} }
  \caption{The contour plots of branching ratios of $B_{d}$ ${\to}$ $K^{+}K^{-}$
  and $B_{s}$ ${\to}$ ${\pi}^{+}{\pi}^{-}$ decays as functions of the
  annihilation parameters ${\rho}_{A}$ and ${\phi}_{A}$ without and with
  the ${\Phi}_{B2}$ contributions in (a) and (b), respectively.
  The solid curves correspond to the central values of data and
  the bands correspond to the $2\,{\sigma}$ constraints.}
  \label{fig-br}
  \end{figure}

  The constraints on annihilation parameters from data are illustrated in Fig.\ref{fig-br}.
  It is clearly seen from Fig.\ref{fig-br}(a) that it is impossible to accommodate simultaneously
   $B_{d}$ ${\to}$ $K^{+}K^{-}$ and $B_{s}$ ${\to}$ ${\pi}^{+}{\pi}^{-}$
  decays within $2\sigma$ errors with the same values of ${\rho}_{A}$
  and ${\phi}_{A}$ when the ${\Phi}_{B2}$
  contributions are overlooked. Other studies of $B$ decays, such as
  Refs.\cite{npb675.333,prd88.014043}, have uncovered similar results.
  It seems not easy to clarify discrepancies between data and the QCDF
  results with the same set of parameters ${\rho}_{A}$ and ${\phi}_{A}$.
  To clam down this situation, the factorizable and nonfactorizable
  annihilation parameters corresponding to different topologies are
  introduced in Refs.\cite{prd90.054019,plb740.56}.
  However, more annihilation parameters make the method uneconomical
  and unsatisfactory.
  Interestingly, by including the ${\Phi}_{B2}$ contributions, Fig.\ref{fig-br}(b)
  shows overlapping areas of annihilation parameters, which implies that the
  ${\Phi}_{B2}$ contributions are nontrivial for accommodating the tension between
  data and QCDF predictions for
  ${\cal B}(B_{d}{\to}K^{+}K^{-})$ and ${\cal B}(B_{s}{\to}{\pi}^{+}{\pi}^{-})$.
  In addition, if theoretical uncertainties from inputs are taken into account,
  the overlapping bands will be inevitably enlarged.
  The same annihilation parameters suitable for pure WA hadronic $B$ decays
  might be obtained with the QCDF approach.

   As is shown by Fig.\ref{fig-br}(b), strict limits on annihilation
   parameters ${\rho}_{A}$ and ${\phi}_{A}$ can not be obtained only from
   experimental data on ${\cal B}(B_{d}{\to}K^{+}K^{-})$ and
   ${\cal B}(B_{s}{\to}{\pi}^{+}{\pi}^{-})$.
   In principle, considering more $B$ decays,  such as a
   global fit on nonleptonic $B$ decays in Refs.\cite{prd90.054019,plb740.56},
   is helpful for extracting the informations of annihilation parameters.
   However, for many hadronic $B$ decays, other contributions, such as
   spectator scattering interactions, will complicate the determination
   of annihilation parameters. How to get annihilation parameter spaces
   as compact as possible from data is beyond the scope of this paper.


   \begin{table}[t]
   \caption{The $CP$-averaged branching ratios in the unit of $10^{-7}$,
   where the theoretical uncertainties are from the input parameters
   listed in Table \ref{tab:inputs}. Different scenarios are explained
   in the text.}
   \label{tab:brs}
   \begin{ruledtabular}
   \begin{tabular}{lcccccc}
         & \multicolumn{4}{c}{our results}
         & Ref. \cite{npb675.333}
         & \\ \cline{2-5} \cline{6-6}
   decay & \multicolumn{2}{c}{S1} 
         & \multicolumn{2}{c}{S2} 
         & S3 & \\ \cline{2-3} \cline{4-5} \cline{6-6}
   mode  & $A_{k}^{i,B2}$ $=$ $0$ & $A_{k}^{i,B2}$ ${\neq}$ $0$
         & $A_{k}^{i,B2}$ $=$ $0$ & $A_{k}^{i,B2}$ ${\neq}$ $0$
         & $A_{k}^{i,B2}$ $=$ $0$ & data \\ \hline
      $B_{s}$ ${\to}$ ${\pi}^{+}{\pi}^{-}$
   &  $3.13^{+0.56}_{-0.43}$
   &  $5.08^{+1.05}_{-0.86}$
   &  $3.44^{+0.62}_{-0.47}$
   &  $5.63^{+1.18}_{-0.97}$
   &  $1.49$ 
   &  $6.7{\pm}0.8$ \\
      $B_{d}$ ${\to}$ $K^{+}K^{-}$
   &  $0.85^{+0.17}_{-0.14}$
   &  $1.01^{+0.20}_{-0.16}$
   &  $0.78^{+0.15}_{-0.13}$
   &  $0.91^{+0.18}_{-0.15}$
   &  $0.79$ 
   &  $0.80{\pm}0.15$
   \end{tabular}
   \end{ruledtabular}
   \end{table}

   \begin{table}[h]
   \caption{The values of $A_{k}^{i}$ in Eq.(\ref{qcdf-p-wa-01}-\ref{qcdf-p-wa-03}).
   See text for explanations of different scenarios.}
   \label{tab:ai}
   \begin{ruledtabular}
   \begin{tabular}{lccccccc}
   decay &  & \multicolumn{3}{c}{$A_{k}^{i,B2}$ $=$ $0$}
            & \multicolumn{3}{c}{$A_{k}^{i,B2}$ ${\neq}$ $0$}
         \\ \cline{3-5} \cline{6-8}
   mode  & scenario & $A_{1}^{i}$ & $A_{2}^{i}$ & $A_{3}^{i}$
           & $A_{1}^{i}$ & $A_{2}^{i}$ & $A_{3}^{i}$ \\ \hline
     $B_{s}$ ${\to}$ ${\pi}^{+}{\pi}^{-}$
   & S1 & $116$ & $116$ & $0$
        & $117$ & $169$ & $-20$ \\
   & S2 & $103-i\,64$ & $103-i\,64$ & $0$
        & $104-i\,67$ & $139-i\,111$ & $-15+i\,17$ \\ \hline
     $B_{d}$ ${\to}$ $K^{+}K^{-}$
   & S1 & $115$ & $115$ & $0$
        & $116$ & $170$ & $-21$ \\
   & S2 & $102-i\,64$ & $102-i\,64$ & $0$
        & $103-i\,67$ & $140-i\,113$ & $-15+i\,18$
   \end{tabular}
   \end{ruledtabular}
   \end{table}

   It is seen from Fig.\ref{fig-br}(b) that, in general, the value of
   ${\rho}_{A}$ increase with the increasing value of ${\vert}{\phi}_{A}{\vert}$.
   A large value of parameter ${\rho}_{A}$ will spoil the self-consistency
   and confidence level of the QCDF approach, and ${\rho}_{A}$ ${\le}$ $1$
   is proposed in Refs.\cite{npb606.245,npb675.333}.
   The strong phase ${\phi}_{A}$ describes the rescattering among hadrons
   and relates closely to $CP$ violation of nonleptonic $B$ decays.
   Focusing on the pure WA decays of $B_{d}$ ${\to}$ $K^{+}K^{-}$ and
   $B_{s}$ ${\to}$ ${\pi}^{+}{\pi}^{-}$, to roughly estimate branching
   ratios, two scenarios based on Fig.\ref{fig-br}(b) are considered in
   our numerical calculation.
   Scenario S1 is with parameters ${\rho}_{A}$ $=$ $1$ and ${\phi}_{A}$
   $=$ $0^{\circ}$, and scenario S2 is with ${\rho}_{A}$ $=$ $1.2$ and
   ${\phi}_{A}$ $=$ $-40^{\circ}$.
   Practically, for the scenario S1, it is intuitive that zero strong phase
   ${\phi}_{A}$ seems a little unnatural.
   Trying to combine the value of ${\rho}_{A}$ as close to one as possible with
   a nonzero ${\phi}_{A}$, the scenario S2 is considered. In addition,
   the scenario S2 is comparable with the scenario S3 of Ref.\cite{npb675.333},
   where the ``universal annihilation'' parameters ${\rho}_{A}$ $=$ $1$
   and ${\phi}_{A}$ $=$ $-45^{\circ}$ are used.

   Using such inputs, we list the QCDF results for ${\cal B}(B_{d}{\to}K^{+}K^{-})$
   and ${\cal B}(B_{s}{\to}{\pi}^{+}{\pi}^{-})$ with and without considering
   the ${\Phi}_{B2}$ contributions  in Table~\ref{tab:brs}, in which
   the theoretical predictions of scenario S3 of Ref.\cite{npb675.333} and
   experimental data are also listed for convenience of comparison.
   In order to show the effects of ${\Phi}_{B2}$ much more clearly,
   we collect  the numerical results of $A_i^k$ in Table~\ref{tab:ai}.

   From Table~\ref{tab:brs}, it can be found that:
   (i) The experimental data for both $B_{d}$ and $B_{s}$ decays can not be well
   explained simultaneously by QCDF approach without considering
   the ${\Phi}_{B2}$ contributions;
   (ii) The numerical difference between the case for $A_{k}^{i,B2}$ $=$ $0$ of
   scenario S2 and scenario S3 of Ref.\cite{npb675.333} arises from different
   inputs, such as decay constants, the CKM parameters and so on, besides
   parameters ${\rho}_{A}$ and ${\phi}_{A}$.
   (iii) With the scenario S2, the ${\Phi}_{B2}$ contributions present
   about $60\%$ and $20\%$ corrections to ${\cal B}(B_{d}{\to}K^{+}K^{-})$ and
   ${\cal B}(B_{s}{\to}{\pi}^{+}{\pi}^{-})$, respectively,
   which significantly improve the QCDF predictions and can explain
   the data within uncertainty.

   The results in Table~\ref{tab:brs} show that ${\Phi}_{B2}$ contributions
   to nonfactorizable WA amplitude building blocks $A^{i}_{k}$ are small,
   due to the small moment ${\langle}{\xi}{\rangle}_{B2}$.
   In addition, according to the conventions of Refs.\cite{npb606.245,npb675.333},
   building block $A^{i}_{3}$ is always accompanied by the small value
   of Wilson coefficient $C_{5}$.
   Hence, on one hand, the dominant contributions to WA amplitudes come
   from ${\Phi}_{B1}$ part; on the other hand, to some certain extent,
   the ${\Phi}_{B2}$ contributions present un-negligible correction to the
   amplitude especially for the pure annihilation decay modes and
   can improve the performances of the QCDF approach.

   In summary, the improvements of measurement precision with the running
   Belle-II and LHCb experiments call for the refinements of theoretical
   calculation on hadronic $B$ weak decays.
   For the $B$ mesons, there are two scalar DAs ${\Phi}_{B1}$ and ${\Phi}_{B2}$.
   The ${\Phi}_{B2}$ contributions to formfactors and branching ratios
   can be significant for some cases with the pQCD approach.
   In this paper, we study the ${\Phi}_{B2}$ contributions with the QCDF
   approach, and find that for the $B$ ${\to}$ $PP$ decays, they can be
   safely neglected in the spectator scattering amplitudes,
   and contribute to only the nonfactorizable WA amplitudes.
   The ${\Phi}_{B2}$ contributions to WA amplitudes are small compared
   with the dominant ${\Phi}_{B1}$ contributions, due to the small
   moment ${\langle}{\xi}{\rangle}_{B2}$.
   However, the participation of ${\Phi}_{B2}$ plays a positive role in
   accommodating the pure WA decays $B_{d}$ ${\to}$ $K^{+}K^{-}$ and
   $B_{s}$ ${\to}$ ${\pi}^{+}{\pi}^{-}$ to data with the universal
   annihilation parameters ${\rho}_{A}$ and ${\phi}_{A}$.
   The values of annihilation parameters ${\rho}_{A}$ and ${\phi}_{A}$
   with the QCDF approach have been under discussion for a long period.
   More information about WA parameters ${\rho}_{A}$ and ${\phi}_{A}$
   could be obtained by a comprehensive study on nonleptonic $B$ decays.

  \section*{Acknowledgments}
  This work is supported by the National Natural Science Foundation of
  China (Grant Nos. 11875122, 11705047, U1632109 and 1191101296) and
  the Program for Innovative Research Team in University of Henan
  Province (Grant No.19IRTSTHN018).

  
  \end{document}